\documentclass{svjour3}
\usepackage{amssymb,amsmath}
\usepackage{hyperref}
\usepackage{graphicx}

\begin{document}

\title{Self-organization of solar magnetic fields}

\titlerunning{Self-organization of solar magnetic fields}

\author{T.R. Jarboe, T.E. Benedett, C.J. Everson, C.J. Hansen, A.C.
Hossack, K.D. Morgan, B.A. Nelson, J.M. Penna, and D.A. Sutherland}

\authorrunning{Jarboe \emph{et al.}}

\institute{University of Washington, Seattle Washington, 98195}

\date{}

\maketitle

\begin{abstract}

Self-organization properties of sustained magnetized plasma are applied to selected solar data to understand solar magnetic fields.
Torsional oscillations are speed-up and slow-down bands of the azimuthal flow that correlate with the solar cycle, and they imply the existence of a symmetric solar dynamo with a measured polar flux of $3 \times 10^{14}$~Wb.
It is shown that the solar dynamo is thin ($\approx$0.1 Mm gradient scale size) and powerful ($\approx$10$^{23}$~W).
These properties are found from the amplitude of the torsional oscillations and the relationship of their velocity contours to solar magnetograms supports the result. 
The dynamo has enough power to heat the chromosphere and to power the corona and the solar wind. 
The dynamo also causes a rigid rotation of the heliosphere out to at least the corona and the relationship of the rotation of the corona to solar magnetograms supports this result as well.
The thin solar dynamo sustains a thin stable minimum energy state that seems to be covering most of the solar surface just below the photosphere. 
The magnetic field lines of the minimum energy state should be parallel to the solar surface and rotate with distance from the surface with $2\pi$  radians of rotation in $\approx$1~Mm. 
Resistive diffusion helps to push the magnetic fields to the surface and the global magnetic structure (GMS) seems to lose $\pi$ radians every 11 years, causing the observed 180$^{\circ}$ flipping of the solar magnetic field. 
The thin sheets of magnetized plasma in solar prominences may be the lost thin sheets of the GMS. 
For completeness, the formation of sunspots, CMEs and flares is discussed. 

\end{abstract}

\keywords{Sun: Magnetic fields, oscillations, rotation, evolution, activity}

\section{Introduction}

\subsection{History}

A century from the first observations (Hale, 1908) of the solar magnetic field, consensus still has not been reached (Charbonneau, 2010) as to the specifics of the mechanism by which that field is generated. 
That should not be terribly surprising, seeing as the problem is one of defining the inner workings of a tremendously complicated system that, until the past few decades, could not be directly sampled past the photosphere.

While George Ellory Hale's pioneering discoveries of the field in sunspots, Joy's Law, and the twenty-two-year cycle, showed that the magnetic field in sunspots were almost certainly due to the existence of an organized magnetic field structure within the Sun, his observations were of the surface only and could not themselves point the way to whatever precise mechanism produced such a field. 
Joseph Larmor (Larmor, 1919) identified flows of conducting fluid, \emph{i.e.}\ the solar plasma, as a possible source for the solar magnetic field: a solar dynamo. 
Larmor's axisymmetric and equatorially antisymmetric model meshed well with Hale's observations.

However, in 1934, Thomas Cowling proved (Cowling, 1934), in what is now known as his antidynamo theorem, that no steady-state axisymmetric field could be produced by an axisymmetric dynamo. 
Thus Larmor's model was mathematically invalid, and consequently the solar dynamo would have to be more complicated than Larmor's, if the solar magnetic field was produced by a dynamo at all. 
Some did doubt the existence of such a dynamo, such as Hannes Alfvén, who in 1942 proposed that the solar magnetic field was simply a remnant from the formation of the solar system (Alfv\'{e}n, 1942).

However, a new instrument --- Horace Babcock's solar magnetograph (Babcock, 1953) --- produced new data, particularly that the entire solar field, not just that in sunspots, reversed every eleven years. 
This rendered Alfvén's model obsolete, and restored the dynamo theory to the lead (Stenflo, 2015), for the obstacle discovered by Cowling had been overcome in the meantime, with Walter Elassier's development of a
consistent dynamo theory, which he applied to the magnetic field of Earth (Elassier, 1946; Elassier, 1955).

Observations and discoveries like these enabled the creation of new dynamo models for the Sun, among them the model proposed by Eugene Parker in 1955, that turbulence plus the Coriolis force could convert toroidal field to poloidal field, while the differential rotation of the Sun plus magnetic buoyancy could convert poloidal field to toroidal field (Parker, 1955). 
This model would come to be systematized into mean-field electrodynamics (Steenbeck \& Krause, 1966), which parameterizes these conversions by decomposing variables into average and fluctuating quantities and solving for the former without directly solving for the latter (R\"{a}dler, 2014), to avoid the problem of solving across the tremendous variation in length scales possessed by a star.

The Parker model was not the only dynamo theory that emerged in this
era; Babcock (Babcock, 1961) and Leighton (Leighton, 1969) developed a
solar dynamo theory of their own, which placed the poloidal field
generation near the solar surface. The toroidal field generation,
however, still was placed deeper in the Sun, which posed the challenge
of how the dynamo could manage to link the two field components'
generation together across the solar convective zone, and this challenge
made this model less popular than its alternative (Tobias, 2002).

Developments since the 1980s posed challenges to all dynamo models then
in existence, from the reevaluation of magnetic buoyancy and diffusivity
effects, to the simulation of existing models of the Sun, to the
development of helioseismology, which allowed the interior of the Sun to
be measured more directly (Charbonneau, 2010). While these developments
have afforded new opportunities (such as the discovery of the tachocline
as a potential location for the solar dynamo), the field of solar
physics has not yet settled into a standard model of the solar dynamo
(Choudhuri, 2007).

Helioseismic measurements have produced (Christensen-Dalsgaard, 2002) an
image of the solar interior with a resolution on the order of 1 Mm
(Hindman, 2004). Features discovered via this technique, such as the
tachocline (Tobias, 2002) and meridional flow have formed key components
of modern solar dynamo theories (Choudhuri, 1995; Haber \emph{et al.},
2002).

If the magnetic field of the entire Sun (or a thick structure within the Sun) is to change in 11 years, it must diffuse through some mechanism that is far faster than resistive diffusion. 
Some proposed mechanisms for solving that problem include ambipolar diffusion or turbulent transport (Parker, 1963). 
However, Parker's later work on the problem says that such mechanisms are not truly applicable to the problem of \emph{vector} diffusion, and these ``[f]undamendal difficulties with the concept of turbulent diffusion of magnetic fields suggest the solar dynamo needs to be reformulated'' as there is ``no way to account for the [standard] value $\eta \approx 10^{12}$~cm$^{2}$/sec, suggesting that it is necessary to rethink the $\alpha \omega$-dynamo for the Sun'' (Parker, 2008). 
Now the situation is not as bleak as Parker describes, with a great deal of research describing the processes which generate this turbulence-driven diffusion (Charbonneau, 2014) and studies of the role of magnetic reconnection activity in the Sun (Yamada, 2010).

\subsection{Requirements of a Global Magnetic Structure model}

Any model of the solar dynamo must therefore be consistent with both the observations made of the Sun and its magnetic field, as well as with the observations made of the behavior of plasmas in terrestrial
laboratories.
The resulting criteria include:

First, the solar magnetic structure is global, encompassing the entire body --- as with Earth, the Sun's overall magnetic field has two opposing magnetic poles, roughly around the poles of rotation --- but only most of the time.

Second, the solar magnetic field reverses itself every eleven years, returning to its original state, more-or-less, after a period of twenty-two years in total (Babcock, 1961).

Third, the solar magnetic field is correlated with sunspot activity, \emph{i.e.}, surface phenomena, which suggests that there should be a close link between the solar surface and the region of magnetic field generation.

Fourth, solar prominences are made of thin sheets of magnetized plasma and the solar corona changes its shape during the solar cycle.

Fifth, torsional oscillations are correlated with magnetic activity (Howe, 2009) and follow the solar cycle. 
They also exist during solar minima.

Sixth, the solar magnetic structure needed to be consistent with present knowledge of plasma self-organization.

\subsection{Proposed Solution}
\label{sec:1.3}

Consideration of the aforementioned constraints from solar observations and experimental exploration into plasma physics suggest that a thin layer near the surface ($ R \approx 0.9986R_{o}$ ) should be considered as the source of the external solar magnetic field.

First, studies of the surface and subsurface of areas from where magnetic structures emerge have shown there is no indication of significant movement to the area before the emergence (Birch \emph{et al.}, 2013), which would suggest either that the structures emerge at truly breakneck speeds such that the motion outpaces the time resolution of the studies, or, more plausibly, that there is very little motion at all, which would indicate that the magnetic field has its source near the surface.

Second, consistent with the picture presented in this paper, the polar flux has a thin cusp source giving the localized corona (Amenomori, Bi, and Chen, 2013).

Third, the recent approximate halving of the dynamo power, in cycle 24, lowers the solar irradiance by the order of 0.02\% (an estimate based on Figure 1 of (Kopp, 2016)). 
Assuming that the change in irradiance and dynamo power are directly connected, then the dynamo power was of order 0.04\% of the irradiance ($1.2 \times 10^{23}$~W of dynamo power with large uncertainly) in 1998 when the signed polar flux was $3 \times 10^{14}$~Wb (Jiang, Cameron, Schmitt, and Schüssler, 2011). 
It will be shown in Section~\ref{sec:3} that the gradient scale length of the magnetic structure was of order 0.1~Mm, a thin structure.

Fourth, rapidly changing magnetic activity in the photosphere is attributed to a ``magnetic carpet'' near the surface (Savage, Steigerwald, and Title 1997). 
These magnetic field may be a hint of something more significant a little deeper.

Fifth, some have argued that torsional oscillations are strongest near the surface (Spruit, 2002) some are less convinced (Howe, 2009). 
A thin shallow dynamo could drive the oscillations strongest near the surface.
The torsional oscillations follow the solar cycle. 
They also exist during solar minima.

The main argument against near-surface dynamos has been that turbulence is too strong to allow for the formation of large-scale magnetic structures. 
It will be shown herein that with a magnetic structure near the minimum energy state, turbulence does not destroy the structure nor helicity. 
A global magnetic structure (GMS) appears to exist in the thin region near the photosphere where the conditions allow magnetic self-organization to have a significant effect on plasma behavior.

\subsection{Very thin and shallow dynamo and GMS are unique results.}

The current leading hypotheses for the nature of the solar dynamo tend to locate the source of dynamo action at the solar tachocline, the sharp boundary between the rigidly-rotating inner sun and the outer 30\% of the sun where the rate of rotation varies with latitude; the sharp gradient in angular velocity at that location is taken as able to advectively stretch and twist and wrap magnetic field lines and so sustain the magnetic structure against Ohmic decay (Charbonneau, 2010; Dikpati and Charbonneau, 2010). 
That is, the most widely-accepted models in the field today place the source of magnetic field generation at a depth of 30\% of the Sun's radius and do not depend on the sustainment of a stable equilibrium to do so.

Historically, no widely-considered model has ever placed the entirety of the solar magnetic field generation very close to the solar surface; the model developed in the 1960s by Babcock (Babcock, 1961) and Leighton (Leighton, 1969) did place some magnetic field generation at the surface, but only for the poloidal field; the toroidal field remained, in their model, generated deeper in the sun, raising questions of how to link the generation of the two field components. 
Alternative models are consistent with the data, such as a dynamo driven by near-surface shear (Brandenburg, 2005) and by a solar dynamo surface wave (Parker, 1993).
However, these model still has the dynamo depth at 35~Mm and 20~Mm, respectively, which is much larger than the $\approx$1~Mm derived in this paper.

By contrast with all, the model proposed in this paper locates the source of all magnetic field generation in a thin layer near the surface, not at the tachocline but in the supergranulation region related to the magnetic carpet 
(Priest, Heyvaerts, and Title, 2002) below the photosphere. 
The structure is thin and close to the surface where the solar dynamo may sustain a stable equilibrium. 
The model can explain butterfly diagrams (Maunder and Maunder, 1904), flipping of the polarity of sunspots (Hale \emph{et al.}, 1919), and the flipping of the polar magnetic field (Babcock, 1961). 
The model also explains a relationship between velocity contours of the torsional oscillations (Howe, 2009) and solar magnetograms (Hathoway, 2010).

Self-organization properties of sustained magnetized plasma are applied to selected solar data for a new understanding of solar magnetic dynamics.
This paper is the first to discuss self-organization of sustained plasma as observed in laboratory plasma and to establish the power and scale size of the solar dynamo from torsional oscillations. 
This forms a basis for interpreting other solar magnetic activities, which are the topics of the rest of the paper. 
It is the first to show that a shallower dynamo $\approx$0.1 Mm thick, can be the source for sufficient torque to explain the torsional oscillations and for sufficient power to heat the chromosphere and corona. 
That self-organization allows this dynamo to produce a GMS that causes the 11-year cycle of magnetic field reversal by moving to the surface, is first shown in this paper. 
The GMS can be the previously unknown source for thin sheets of plasma for prominences and for the magnetic flux in sunspots. For this interpretations the Sweet/Parker time is less than a day. Thus, this is the first paper where fast reconnection is not required for solar activity. 
In summary, the application of plasma self-organization knowledge, obtained from studying laboratory plasma, represents a breakthrough in understanding solar electromagnetic activity.

\section{Derivations and observations of the self-organization of magnetic structures}
\label{sec:2}

\subsection{The minimum energy principle}
\label{the-minimum-energy-principle}

In the solar plasma just below the photosphere, the collision frequency is higher than the cyclotron frequency, eliminating the Hall effect.
Therefore, it is well described by resistive magnetohydrodynamics with a generalized Ohm's law, which is found by a Lorentz transformation from the plasma frame, where $\vec{E} = \eta \vec{j}$, to the lab frame.
That generalized Ohm's law is:

\begin{equation}
\vec{E} = -\vec{v \times B} + \eta \vec{j}
\label{eq:1}
\end{equation}


\noindent
where $\vec{E}$ is the electric field, $\vec{v}$ is the plasma velocity, $\vec{B}$ is the magnetic field, $\eta$ is the resistivity, and $\vec{j}$ is the current density. 
This Ohm's law is valid for several megameters below the photosphere. 
For the corona and above the Hall terms are required.

The Woltjer-Taylor minimum energy principle states that magnetic plasmas relax toward a state of \emph{m}inimum \emph{e}nergy while \emph{c}onserving \emph{h}elicity (the MECH state). 
Magnetic helicity is the linkage of magnetic flux with magnetic flux (Moffat, 1978), $ K = \int \vec{A \cdot B}\, d\text{Vol}$, where $\vec{A}$ is the vector potential and $\vec{B}$ is the magnetic field. 
For a closed conducting boundary, with $\vec{B \cdot n}$ = 0 on the boundary, the solution is $\vec{\nabla \times B} = \lambda_{eq} \vec{B}$, where $\lambda_{eq}$  is a global constant, and is called the ``Taylor state'' (Woltjer, 1958; Taylor, 1986). 
Complete relaxation, with magnetic flux boundary conditions where a vacuum helicity is added when static magnetic field pierces the conducting boundary to make the total helicity gauge invariant, also gives $\vec{\nabla \times B} = \lambda_{eq} \vec{B}$, where $\lambda_{eq}$ is a global constant. (Finn and Antonsen, 1985)

MECH magnetic structures are stable and may be tolerant to velocity turbulence. 
Helicity is dissipated by $2\vec{E \cdot B}$ (Finn and Antonsen, 1985). 
While velocity turbulence cannot dissipate helicity (as the velocity-dependent term in $\vec{E}$ is in a term perpendicular to $\vec{B}$  due to the cross product,) it can dissipate excess magnetic energy by $\vec{j \cdot E}$ , which may be a more rapid process than helicity dissipation because the velocity term can dissipate energy. 
(Random turbulence causes a net dissipation of excess energy because velocities that absorb excess magnetic energy increase in time, increasing dissipation, and velocities that increase the magnetic energy decrease in time, decreasing energy input.) 
This would drive the magnetic structure to the MECH state. 
Thus, turbulence dissipates only the excess magnetic energy but does not affect magnetic helicity. 
The magnetic structure can exist in the MECH state and perhaps only in or near the MECH state in the presence of strong turbulence. 
Furthermore, helicity decays on the resistive time scale (the longest characteristic time scale in the system (Edenstraser and Kassab, 1995)) and the MECH state is stable (Edenstrasser and Kassab, 1995).

Utilizing Ampere's law yields $\lambda_{eq} = \mu_{0} j_{\parallel} / B$, where $ \lambda_{eq}$ is the inverse magnetic gradient length.
The dissipation of energy per unit of helicity dissipation is $\left( \vec{j \cdot E} \right) / \left( 2 \vec{E \cdot B} \right) = j/2B = \lambda /2 \mu_0$ so $\lambda$ is proportional to the energy per unit helicity.
Also it is, of course, proportional to the current per unit flux.
For similarly-shaped MECH states, $\lambda_{eq}$ is proportional to 1/size.
The larger class of force-free states obey $\vec{\nabla \times B} = \lambda \vec{B}$ where $\lambda$ is a function of flux (and no subscript is used) but is still the rate of rotation of the magnetic field in the direction across the magnetic field.

Current tends to follow field lines.
(This rule is used in estimating current paths.)
Using Eq.~\ref{eq:1} and the momentum equation $\rho d \vec{v}/dt = \vec{j} \times \textbf{B}$ it is easy to show that an electric field $E_{x}$ :

\begin{equation}
E_x = \eta j_x \left( 1 + S \frac{t}{\tau_\text{Alf}} \right)
\label{eq:2}
\end{equation}

\noindent
is required to drive a current $j_x$ in the $x$-direction, when the initial $\vec{v} \times \vec{B}$ is zero, where $\vec{B}$ is in the $z$-direction and $S$ is the Lundquist number. 
Here $t$ is the time when $E_x$ is applied for a time in the range of $\tau_\text{Alf} < t < \tau_\text{L/R}$.  For times less than the Alfv\'{e}n time, $\tau_\text{Alf}$, $t$ is the Alfv\'{e}n time and for times greater than the resistive diffusion time, $\tau_\text{L/R}$, $t$ is the resistive diffusion time.
Both $\tau_\text{Alf}$ and $\tau_\text{L/R}$ are based on the size 1/$\lambda$. Thus, when the initial $\vec{v \times B}$ is zero, the current tends to follow field lines since the Lundquist number $S$ is large ($S^{2} \approx 10^{15}$).

\subsection{Sustaining stable plasma currents by helicity injection current drive}

Experimental results show that sustained MECH states can be modeled as a two-step $\lambda$ profile (B. Hudson, 2008; Jarboe, 2012). 
Sustained plasmas can have an arbitrary $\lambda$ value as part of the boundary conditions (defined by an external ``injector'' circuit), and the MECH state does not have uniform $\lambda$ (Jarboe, 1994). 
Ideally the injector fields will have much of its boundary parallel to the rest of the equilibrium and a separatrix can form between the region with $\lambda_{\rm inj}$ defined by the boundary conditions and the region where $\lambda_{\rm self}$ is found by this self-organization. 
The final ingredient is perturbations that cause an anomalous viscosity in the electron fluid giving the cross-field current drive from the driven region to sustain the self-organized region resulting in the sustainment of the GMS (Jarboe, 2015; Hossack, 2017).
The perturbations also break the symmetry of the magnetic axis and the sustainment is consistent with Cowling's Theorem. (Cowling, 1934) 
The perturbations can be imposed by the external geometry or circuit or produced by plasma instability. Since the energy per unit helicity is $\lambda / 2 \mu_{0}$ helicity flows from high-$\lambda$ to lower-$\lambda$ regions of the GMS. 
The helicity injection circuit maintains the higher $\lambda_{\rm inj}$. 
This leads to a two-step-like $ \lambda$ profile as observed in the Helicity Injected Torus with Steady Inductive helicity injection (HIT-SI) experiment (Jarboe \emph{et al.}, 2012; Victor \emph{et al.}, 2014).

\subsection{Imposed dynamo current drive}

\begin{figure}
\begin{center}
\includegraphics[width=\linewidth]{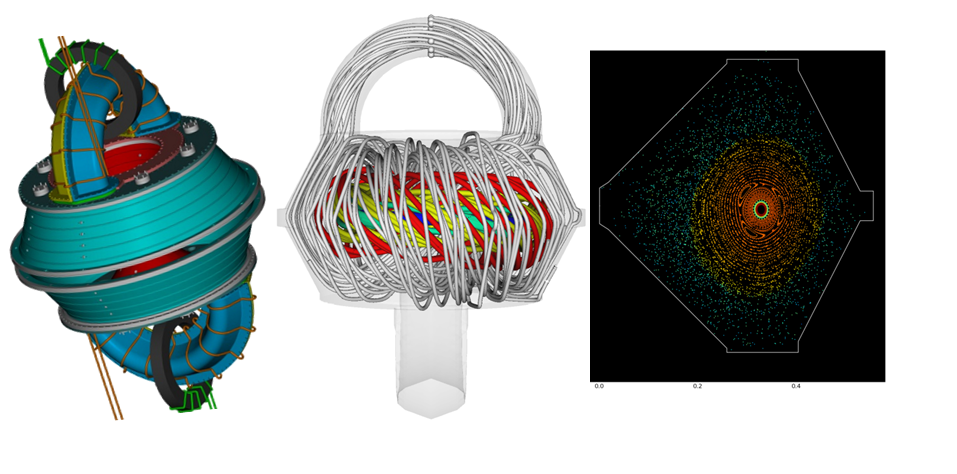}
\end{center}
\caption{Left: a drawing of the HIT-SI device exterior. 
Middle: HIT-SI Taylor state equilibrium magnetic field lines when the top injector is at peak current and loop voltage. 
The gray edge field lines link the injector. 
The colored field lines are on different nested flux surfaces inside the separatrix. $I_\text{tor} / I_\text{inj}$ = 6.
Right: a puncture plot of that equilibrium.}
\label{fig:1}
\end{figure}

There are two requirements for imposed dynamo current drive (IDCD): (1)
Ex\-ter\-nal\-ly-driven edge electron current must have flow speeds higher
than in the dynamo-driven region (injecting helicity with
$\lambda_\text{inj} > \lambda_\text{self}$), and (2) magnetic perturbations must be imposed across the entire plasma cross-section that are sufficiently large to drive the stable current profile.
Figure~\ref{fig:1} shows HIT-SI, the first experiment to meet these requirements. 
HIT-SI consists of an axisymmetric spheromak containment flux conserver and two inductive injectors (x-injector and y-injector) mounted on each end as shown in Fig.~\ref{fig:1}. (Jarboe, 2011) 
The voltage, current, and axial flux of each injector are oscillated in phase in the range of 5.8 kHz to 68.5 kHz. 
The two injectors are 90 degrees out of phase. 
The injectors are purely inductive 180$^\circ$ segments of a toroidal pinch attached to a slotted flux conserver. 
The power and helicity injected by the sum of the two injectors is approximately constant in time during the discharge and a slowly varying, compared to the injector frequency, spheromak is formed and sustained. 
The second frame of Fig.~\ref{fig:1} shows how the driven fields connect to and drive high currents in the edge. 
Because the injectors have $n$=1 symmetry they also impose the perturbations required, eliminating the need for instability. 
The third frame shows closed-flux surfaces of an equilibrium with an $n$=1 distortion.

\subsection{HIT-SI results:}\label{hit-si-results}

Toroidal currents up to 90 kA and up to 3.8 times the quadrature sum of the injector currents are achieved using IDCD on HIT-SI. 
Langmuir probe temperature measurements indicate $T_e$ of 6 eV to 20 eV.
Ion Doppler spectroscopy measurements of carbon lines indicate $T_i$ of 20 eV to 40 eV. 
Fitting the Grad-Shafranov equilibrium to the injection-cycle averaged internal magnetic probe data shows a flat $\lambda$-profile that is stable to the confinement-destroying long wavelength kink modes and the sustained equilibrium is observed to be stable.

Figure~\ref{eq:2} shows the data from a high-current 14.5~kHz and a high-current-gain 68.5~kHz discharge. 
The discharge time is limited because of overheating of the wall and excess density late in time.  
None-the-less, the pulse lengths are much longer than an injector period showing that the method is steady state. 
HIT-SI has run at 5.8~kHz, 14.5~kHz, 36.8~kHz, 53.5~kHz, and 68.5~kHz. 
All data at gains greater than 2.5 agree well with the two-step $\lambda \left( \equiv \mu_{0}j_{\|}/B  \right)$ profile defined in References (Jarboe \emph{et al.}, 2012 and Victor \emph{et al.}, 2014).

\begin{figure}
\begin{center}
\includegraphics[width=\linewidth]{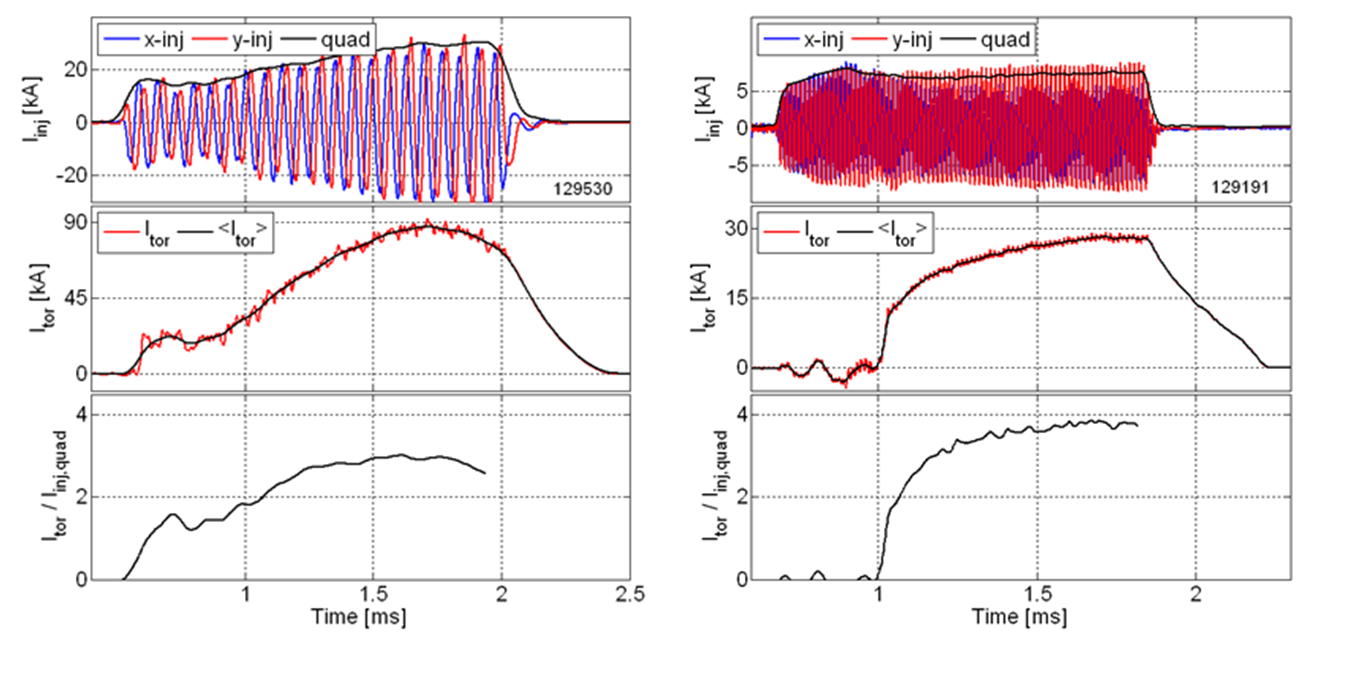}
\end{center}
\caption{Injector current, toroidal current, and their ratio for a 90~kA discharge at 14.5~kHz and for a discharge at 68.5~kHz with a current gain approaching 4.}
\label{fig:2}
\end{figure}

If the Lundquist number defined as the resistive diffusion time normalized by the toroidal Alfv\'{e}n time it is the same order in HIT-SI as in the Sun. 
This Lundquist number characterizes resistive MHD in toroidal geometry. The measurements and simulations of HIT-SI to understand plasma self-organization plus the nature of the solar helicity injector, the solar dynamo, give a clear picture of the sustainment of a solar GMS.

\subsection{Summary of self-organization rules used to understand the solar magnetic dynamics.}

Some useful conclusions from Section~\ref{sec:2} are summarized. 
Magnetic helicity is the linkage of magnetic flux with magnetic flux and magnetic plasma relax toward a state of \emph{m}inimum \emph{e}nergy while \emph{c}onserving \emph{h}elicity (the MECH state). 
For a closed conducting boundary, with $\vec{B \cdot n} = 0$ on the boundary, the MECH state has $\vec{\nabla \times B} = \lambda_{eq} \vec{B}$, where $\lambda_{eq}$ is a global constant. 
MECH magnetic structures are stable and may be tolerant to velocity turbulence.
$\lambda_{eq}$ (=$\mu_0 j / B$) is the inverse magnetic gradient length. 
In plasmas with zero vacuum magnetic field, the energy per unit of helicity is $\lambda_{eq} / 2 \mu_{0}$. 
Of course, $\lambda_{eq}$ is proportional to the current per unit flux and the rate of rotation of the magnetic field in the direction across the magnetic field. 
For similar shaped MECH states $\lambda_{eq}$ is proportional to 1/size. 
Sustained MECH states tend to have a two-step $\lambda$ profile. 
There are two requirements for imposed dynamo current drive (IDCD):
(1) Externally-driven edge electron current must have flow speeds higher than in the dynamo-driven region (injecting helicity with $\lambda_\text{inj} > \lambda_\text{self}$), and
(2) magnetic perturbations must be imposed across the entire plasma cross-section that are sufficiently large to drive the stable current profile. 
These self-organization properties are observed on HIT-SI. 
The solar plasma just below the photosphere, is well described by resistive magnetohydrodynamics. 
Finally, current tends to follow field lines.

\section{The solar dynamo:}
\label{sec:3}

It is shown that the torsional oscillations appear to be caused by a symmetric solar dynamo of uniform $j / B$ and a measured signed polar flux, and that the solar dynamo is thin ($\approx$0.1~Mm gradient scale size) and powerful ($\approx$10$^{23}$~W). 
The solar magnetic activity and the torsional oscillations are well correlated (Howe, 2009) and both are very likely powered by the solar dynamo. 
Therefore, the solar dynamo is assumed to cause the torsional oscillations. 
The solar dynamo is assumed to be dominated by toroidal symmetry because the torsional oscillations are
symmetric, the Sun is symmetric and at a given latitude the probability
of a magnetic event does not appear to depend on longitude. 
Uniform $\lambda_\text{inj}$ is a good approximation in self-organized plasma like the solar plasma as discussed in Section~\ref{sec:2}. 
The measured unsigned flux for 1997 is given in Reference (Jiang, Cameron, Schmitt, and Sch\"{u}ssler, 2011). 
The polar flux is one-half the unsigned flux.

\begin{figure}
\begin{center}
\includegraphics[width=\linewidth]{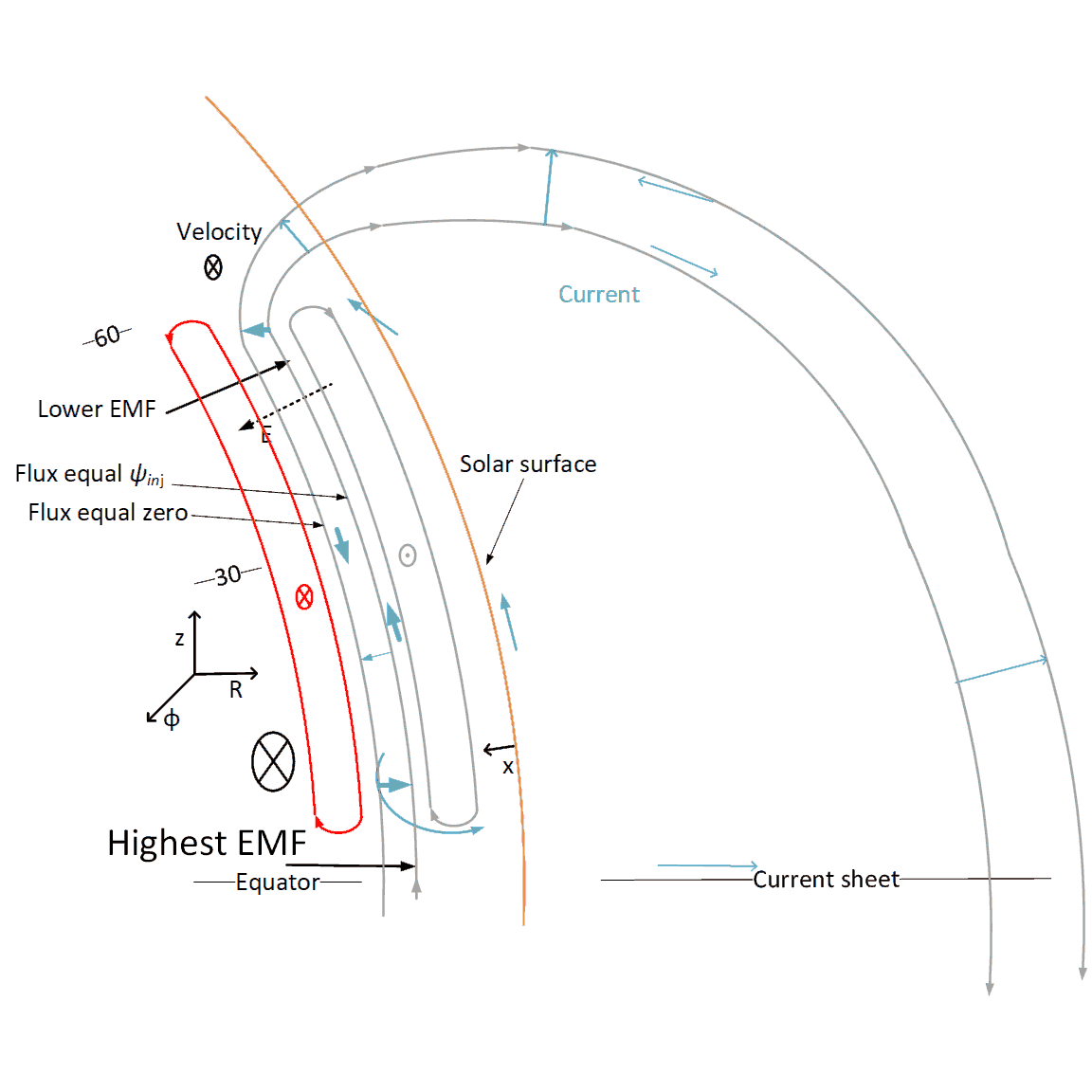}
\end{center}
\caption{Sketch of the expected sustained MECH state. 
The figure represents a snapshot in time of one-half of poloidal cross sections of the dynamo and the closed structures produced by self-organization called “lobes”. 
The red and gray arrows and direction symbols are that of magnetic field. 
The thickness of the dynamo region under the solar surface is much thinner than that shown and the size of the return flux region on the right is much larger than shown. 
Current directions are shown as blue arrows. 
Black arrows are electric field (dashed) and EMF (solid).
The black direction X is for the surface velocity and larger size means higher velocity. 
The polar flux is between the two gray curves. Figure is for the year 1997 (solar minimum).}
\label{fig:3}
\end{figure}

The solar dynamo exists where the open polar flux passes through the sun in Fig.~\ref{fig:3}. 
Using Eq.~\ref{eq:1} and Fig.~\ref{fig:3} and a non-rotating rest frame at the center of the Sun, the EMF voltage is $\int \! v B\, dR$ and the flux, $\psi_{\rm inj}$, is $2\pi R_\text{axis} \int \! B\, dR$. Thus:

\begin{equation}
\text{EMF} = \frac{\omega}{2 \pi} \psi_\text{inj}
\label{eq:3}
\end{equation}


\noindent
where $ R_{\text{axis}}$ is the distance from the surface to the axis of rotation, $\psi_\text{inj}$ is the polar flux that penetrates the solar surface, $v$ is the plasma velocity, and $\omega$ is $v/R_\text{axis}$. 
Thus, the voltage source is the same as in many earlier solar dynamo models.

Assuming $\lambda_\text{inj}$ is uniform the torque about the solar axis is $\vec{\Gamma} = R_\text{axis} \int \vec{I}_\text{cross} \times \vec{B}\, dR$ where $\vec{I}_\text{cross}$ is the current crossing the flux surface at $R$ and equals $\lambda_\text{inj} \psi_\text{inj} / \mu_0$ at $R = R_\odot$ and equals zero on the inside edge of the polar flux, so $I_\text{cross} = \lambda_\text{inj} \psi^{\prime} / \mu_{0}$ where $\psi^{\prime}$ is the flux as a function of $R$. 
Now $\Gamma = \int \lambda_\text{inj} \psi^{\prime} d \psi^{\prime} / 2 \pi \mu_0$,
(the integral is from zero to $\psi_\text{inj}$) or:

\begin{equation}
\Gamma = \frac{\psi_\text{inj} I_\text{inj}}{4 \pi} = \frac{\lambda_\text{inj} \psi^2_\text{inj}}{4 \pi \mu_0} .
\label{eq:4}
\end{equation}

The amplitude of the torsional oscillations is used to estimate $I_\text{inj}$. 
Torsional oscillations are speed-up and slow-down bands of the azimuthal flow that correlated with the solar
cycle (Howe, 2009). 
The bands have a depth of $0.1 R_{\odot}$ and solar activity is in the bands with positive acceleration. 
Assume that thermal convection distributes the torque from the dynamo down to the band depth because the convective turnover times are much shorter than the 11-year period (Landin, 2010).
The band flow speeds are consistent with the torque produced by the thin dynamo. Since the torsional oscillations are well correlated to the dynamo activity, assume that the torque of the dynamo is transmitted to drive the torsional oscillations. 
This torque is equal to $\mathcal{I}_{\,t.o.} \alpha_{t.o.}$ where $\mathcal{I}_{\,t.o.}$ is the moment of inertia of the torsional oscillations and $\alpha_{t.o.}$ is their angular acceleration. 
Setting this torque equal to the dynamo torque gives the dynamo current. 
Then using $\mu_0 I_\text{inj} / \psi_\text{inj} = \lambda_\text{inj}$ the inverse gradiant scale length of the dynamo is found. 
The large value of $\lambda_\text{inj}$ shows the dynamo is thin.

\begin{figure}
\begin{center}
\includegraphics[width=\linewidth]{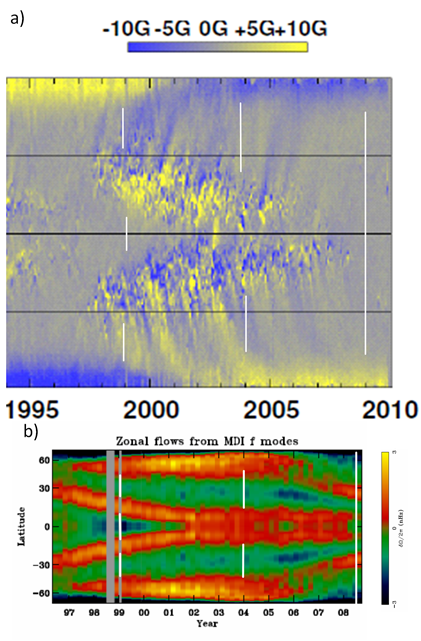}
\end{center}
\caption{a) ``A Magnetic Butterfly Diagram constructed from the longitudinally averaged radial magnetic field obtained from instruments on Kitt Peak and SOHO. 
This illustrates Hale’s Polarity Laws, Joy’s Law, polar field reversals, and the transport of higher latitude magnetic field elements toward the poles.'' (Hathaway, 2010). 
White lines added to show position of dynamo under the solar surface.
b) The contours of the solar rotation frequency change due to the torsional oscillations from (Howe, 2009) with permission of Jesper Schou.}
\label{fig:4}
\end{figure}

Other solar data supports the thinness of the dynamo. 
Torsional oscillations occur in the region of 0.9 solar radii and above. 
Estimate $\mathcal{I}_{\,t.o.}$ for one hemisphere as a point mass of three-fourths the mass above $R = 0.9 R_{\odot}$ and located at a radius of $0.9 R_\odot$. 
The mass per unit area is the pressure at $0.9 R_\odot$ divided by $g$ (the gravitational acceleration, assumed to be 275 m s$^{-2}$) yielding $7.64 \times 10^{8}$~kg m$^{-2}$ and a simple calculation gives $\mathcal{I}_{\,t.o.} = 5.7 \times 10^{44}$~ kg m$^{2}$. 
For $\alpha_{t.o.}$ estimate a 6~nHz change in rotation frequency in 11 years since this is about the peak to peak amplitude of the torsional oscillations (Fig.~\ref{fig:4}b). 
This yields $10^{-16}$~s$^{-2}$. 
The torque is $6.18 \times 10^{28}$~N m.
The signed polar flux at solar minima in 1997 was $3 \times 10^{14}$~Wb (Jiang, Cameron, Schmitt, and Schüssler, 2011). 
This gives $I_\text{inj} = 2.58 \times 10^{15}$~A. 
This gives $\lambda_\text{inj} = 10^{-5}$~m$^{-1}$ or gradient scale length of 100 km.

A dynamo of this size can power the chromosphere, the corona, and the solar wind. 
The highest angular velocity on the solar surface is at the equator, with the angular velocity is almost uniform from 0 to 15$^{\circ}$ latitude (Christensen-Dalsgaard and Thompson, 2007). 
At the equator, the rotational frequency is given by $\omega / 2 \pi \approx 460$~nHz. The
frequency difference between the equator and 60$^{\circ}$ latitude is 90 nHz. This give the voltage across the polar flux of 138~MV and across the highest flux surface of 27~MV. 
Thus the power available for solar activity across the highest flux surface is $7 \times 10^{22}$~W per hemisphere which is enough to power the chromosphere, the corona, and the solar wind. 
Also it is about one-half the measured total power from magnetic activity discussed in Section~\ref{sec:1.3}. 
(It is believed that $2.65 \times 10^{22}$~W are needed to heat the whole chromosphere (Vernazza, Avrett, and Loeser, 1981). 
To heat the quiet corona takes about $2 \times 10^{21}$~W, and active areas have 30 times the quiet power per unit area (Klimchuk, 2006). 
Conservatively assuming 10\% active area, the required power is $8 \times 10^{21}$~W.)

Dynamo action also occurs during times of solar activity. 
Anytime flux enters and leaves the sun at locations of different rotational rates, dynamo drive is produced. 
The crossings with the dynamo drive are below the surface and take energy out of the rotation. 
Thus, the gray areas in Fig.~\ref{fig:4}a) have negative acceleration in Fig.~\ref{fig:4}b). 
White lines mark the dynamo driven region at three different times.
Note the two single-hemisphere dynamos of Fig.~\ref{fig:4} (around the year 2005) are shown by the gray areas in Fig.~\ref{fig:4}a) and negative acceleration in Fig.~\ref{fig:4}b). 
The current crossing in the direction of the electric field are giving power to the heliosphere where field lines penetrate the surface, resulting in positive acceleration. 
Thus the blue and yellow areas of Fig.~\ref{fig:4}a) have positive acceleration in Fig.~\ref{fig:4}b).

The point of highest EMF on a flux surface has current flowing through it in the direction of the EMF and drives current against the EMF at places on the flux surfaces with lower EMF. 
In Fig.~\ref{fig:3}, from the highest flux surface, $\psi_\text{inj}$, the current can flow across this flux surface to raise any lower voltage area. 
In particular, the places under the surface where the EMF is lower and the density is high with convection distributing the force over a large volume (inhibiting acceleration and preventing a back EMF), makes a low
impedance path that takes most of the dynamo current. 
Where the polar flux expands out of the GMS would be such a place. 
The thick current arrows show this path. 
The torque from this current drives the torsional oscillations.

The solar dynamo forces a rigid motion of the heliosphere out to the corona. 
The heliosphere is much lower density then under the solar surface. 
Where polar flux exits the photosphere, the powerful dynamo torque quickly accelerates the heliosphere until the back EMF equals the dynamo driven EMF on each flux surface. 
Since the flux is the same, forcing the EMF to be the same, forces the rotation rate, $\omega$, to equal that of the solar surface (see Eq.~\ref{eq:3}).

\begin{figure}
\begin{center}
\includegraphics[width=\linewidth]{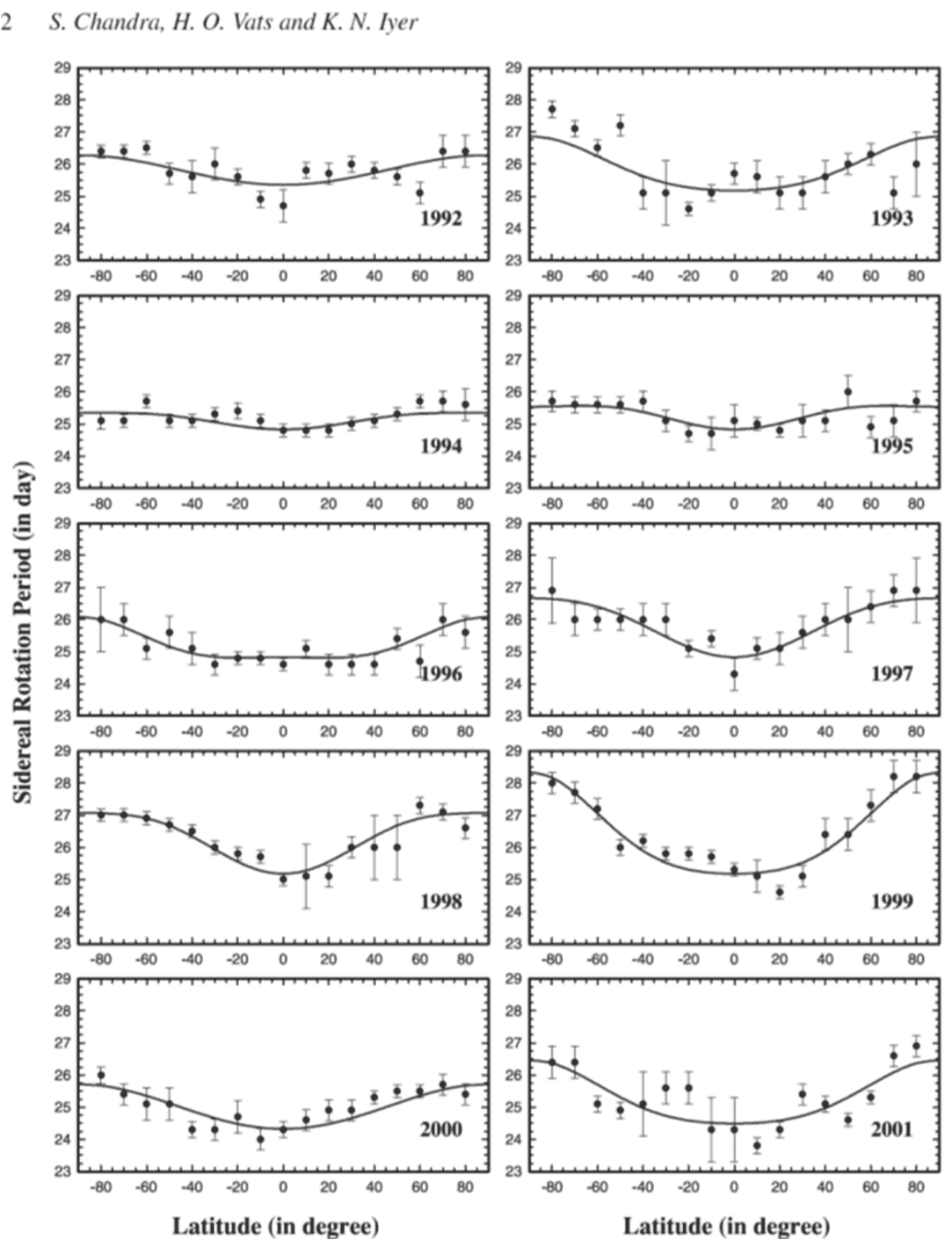}
\end{center}
\caption{Rotation times of the corona versus latitude for different years as measured with the flux modulation method. From (Chandra, 2010).}
\label{fig:5}
\end{figure}

Studying soft-x-rays from the edge of the corona (Chandra, 2010) confirms details of the dynamo driving the rotation of the heliosphere.
For example, in 1999 (Fig.~\ref{fig:4}a) the equator driven dynamo is limited in latitude and does not dominate the entire Sun as it does in 1996--7. 
Thus in 1999 the upper-latitude rotation of the corona is set more by the upper-latitude surface rotation and is slower. 
The white lines define the three dynamo regions in 1999. 
See Fig.~\ref{fig:5}, which is from Reference (Chandra, 2010). 
Also, in Fig.~\ref{fig:5} observe the very uniform rotation in 1994. 
At this time there are two dynamos in Fig.~\ref{fig:4} with the same maximum EMF giving more control of the rotation. 
Thus at this time the corona has a more uniform rotation rate than at solar minimum.

The value of $I_\text{inj}$ is consistent with the value of the current sheet current measured at the earth. 
In the GMS the injector current is parallel to the injector flux with $\lambda_\text{inj} = \mu_0  I_\text{inj} / \psi_\text{inj}$. 
When the injector flux leaves the GMS it expands with distance from the GMS but the total polar flux into an expansion surface is conserved. 
Since $\lambda \approx$ 1/size, $\lambda$ drops from a decrease in the parallel current, $I_{\|}$, so:

\begin{equation}
\frac{I_{\|}}{\lambda} = \frac{I_\text{inj}}{\lambda_\text{inj}} .
\label{eq:5}
\end{equation}


$I_{\|}$ is the current flow parallel to the polar flux into the surface of expansion. 
Kirchhoff's law requires the current return and this appears to be the current sheet. 
At Earth, the current sheet was measured (Smith, 2001) and has 9~GA. 
Using $\lambda$ = 3/(radius of earth's orbit) Eq.~\ref{eq:5} gives agreement with 9.6~GA. 
$I_{\|}$ decreases with expansion because current must cross the magnetic field in the process of converting magnetic energy to plasma kinetic energy.


\section{Magnetic activity generated by the thin dynamo through self-organization}

While it has been shown that the dynamo is thin, it is still necessary to give plausibility arguments of how such a dynamo can produce the observed solar magnetic activity. 
While the descriptions to be given are plausible ways that the various phenomenon can be produced, these are
basically volumetric behaviors that produce surface data and other plausible explanations for the data are possible. 
The self-organized properties of magnetized plasma given in Section~\ref{sec:2} are used. 
A computer code is being developed to explore simulating this plasma.

\subsection{Resistive diffusion leading to surface magnetic activity, prominences and the 11 year cycle:}

A thin, shallow GMS will be explored for the following reasons: the gradient scale size of the solar dynamo as shown in Section~\ref{sec:3} is $\approx$0.1~Mm and such short scale sizes of other quantities exist near the solar surface; magnetic carpet data suggest the existence of magnetic structures just below the photosphere that may have short gradient scale size (Priest, 2002); and self-organization physics allows such a structure. 
Helicity injection (Taylor, 1986) from the solar dynamo, as discussed in Section~\ref{sec:2}, sustains the GMS, and instability might provide the perturbations needed for sustainment and for keeping it in the MECH state as discussed in Section~\ref{sec:2}.

The GMS appears to consist of two thin ($\approx$1~Mm thick), sustained MECH states, near the surface of the Sun. Each covers the Sun from 15 degrees, or less, to about 60 degrees latitude with large-scale longitudinal symmetry. 
One MECH state is in the northern hemisphere and is coupled by common polar magnetic flux to one in the southern hemisphere. 
In Fig.~\ref{fig:3}, the GMS ends where the polar flux exits the self-organized lobed structures. 
If the GMS exists as MECH states, then $\vec{\nabla \times B} = \lambda \vec{B}$  is dominated by the rapid spatial variation in the radial direction, and variations in the other directions can be ignored. 
Thus, if the GMS is in a thin layer it will be approximately a sustained MECH state with the only
relevant variation being in one dimension. 
Solving $\vec{\nabla \times B} = \lambda \vec{B}$  gives the 1D MECH state with the longitudinal and latitudinal magnetic field as $B_\text{long} = -B_{o} \sin \left( \lambda \Delta R \right)$, $B_\text{lat} = B_{o} \cos \left( \lambda \Delta R \right)$, where $ \Delta R$ is the radial position and is zero at the inside edge of the polar flux (also referred to as injector flux).
This sustained MECH state is approximated as a two-step profile with $\lambda_\text{inj}$ in dynamo-driven polar flux and $\lambda_\text{self}$ for the rest of the GMS, which is formed by self-organization. 
This two-step profile is used because it is simple and fits data in driven self-organized plasmas well as discussed in Section~\ref{sec:2}.
The polar flux threads between two lobes of approximately $\pi$ rotation each of the magnetic field as shown in Fig.~\ref{fig:3}. 
Arrows along streamlines and red and gray into the page and out-of the page directions are for the $\vec{B}$-field.
The closed structures are toroidally symmetric lobes of spatially-rotating magnetic field with purely toroidal field (toroidal field is longitudinal field) at the center and mostly poloidal field (poloidal field is perpendicular to the toroidal direction) at the edges.
The northern hemisphere has negative helicity ($\vec{j}$ anti-parallel to $\vec{B}$) and negative $\lambda$, and the southern hemisphere has positive helicity and positive $\lambda$.
Every 11 years $\pi$  radians of rotation escape, causing the observed magnetic activity with the magnetic fields flipping 180$^{\circ}$.
The driven polar flux can sustain the GMS as discussed in Section~\ref{sec:2}.

Our present understanding of self-organization leaves little doubt that a thin GMS will have the MECH state structure and that 180$^{\circ}$ flipping of the solar magnetic field and the surface magnetic fields comes from losing $\pi$ radians of the structure each 11 years.

The details of the GMS structure and the details of the loss of the upper lobe on the 11-year cycle~are not known and more data are needed for a quantitative picture. 
The size, depth, and period of the GMS can be defined by resistivity as is now discussed. 
The following assumptions are made:

\begin{enumerate}
\item The lobe boundaries move by resistive diffusion. 
The unmagnetized plasma outside the GMS is not evolving since self-organization requires magnetic fields. 
Thus the unmagnetized plasma is a resistive flux conserver that confines the GMS. As the magnetic fields diffuse into the boundary the new volume becomes part of the GMS.

\item  The GMS is a 1D MECH state stabilized by local instability and/or high turbulence keeping it near the stable MECH state as discussed in Section~\ref{sec:2}.

\item  Dynamo flux, with $\lambda_\text{inj}$, threads two resistively-expanding lobed structures and gives them helicity injection. 
Two lobes are required so that helicity can be injected without changing the global toroidal flux or global current. 
The dynamo flux must pass between the lobes, as in Fig.~\ref{fig:3}, so that the dynamo flux links the flux of the upper lobe, giving the maximum helicity.

\item  The plasma meets the requirements for sustained two-step MECH state as discussed in Section~\ref{sec:2}. 
The magnetic field rotates $2 \pi$ across the GMS with most of the rotation in the self-organized regions. 
The rotation is $2\pi$ so that the image current stay in the GMS.

\item  Relaxation fills the resistively-produced new volume with GMS, keeping the boundary thin. 
The boundary of thickness $\delta \approx 1/ \lambda_{\rm inj}$ separates changes in $B$, which preserves the ($p + B^2/2\mu_{0}$) jump condition.
The force to confine the magnetic field, as the magnetic field goes to zero at the boundary, is supplied by $\vec{j \times B}$ in the boundary and transmitted to the unmagnetized plasma as pressure. 
This is the force that drives the torsional oscillations. 
From Eq.~\ref{eq:1}, with $\vec{E}$ equal zero, and Amp\`{e}re's Law, the boundaries between the GMS and the unmagnetized plasma then move with velocity:

\end{enumerate}

\begin{equation}
v = \frac{\eta}{\mu_0 \delta}
\label{eq:6}
\end{equation}


Resistivity decreases with depth into the Sun so the top surface moves toward the solar surface faster than the bottom surface moves away, giving a total motion up, towards the solar surface. 
The resistive expansion might be balanced by loss of volume when plasma escapes, moving the GMS upper boundary down to the top of the lower lobe. 
The GMS is stable and its sheared magnetic field has a stabilizing effect on the boundary, which seems to be stable. 
The boundaries move by Eq.~\ref{eq:6} and the GMS does not need neutral buoyancy.

\begin{figure}
\begin{center}
\includegraphics[width=\linewidth]{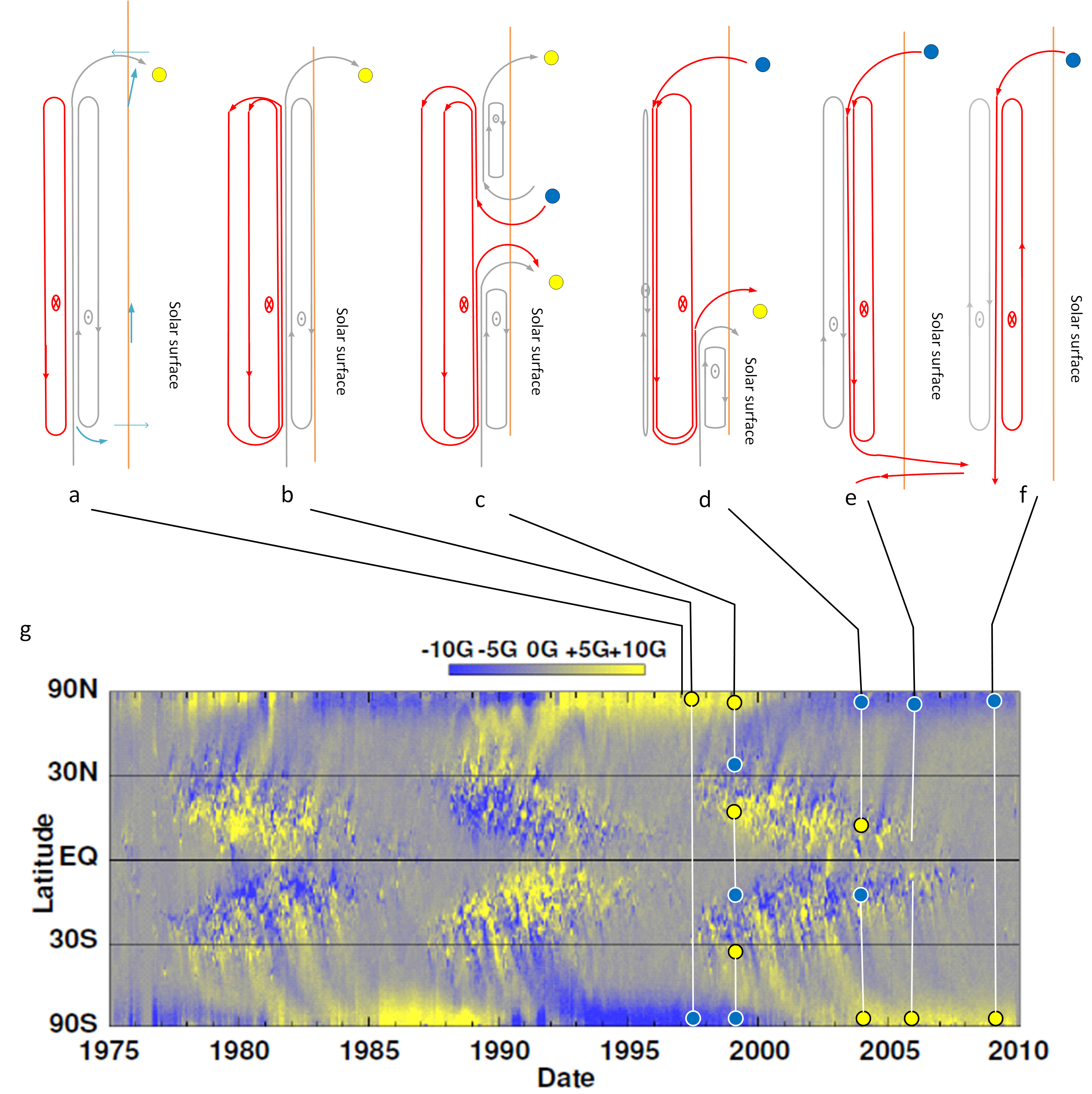}
\end{center}
\caption{(a -- f) Sketch of the process of the polar flux reversal for a single hemisphere only, displaying the formation of new lobes, the displacement of old lobes, and the resistive process by which the polarity reverses. 
The solar surface is sketched as straight for illustration purposes.
g) ``A Magnetic Butterfly Diagram constructed from the longitudinally averaged radial magnetic field obtained from instruments on Kitt Peak and SOHO.
This illustrates Hale's Polarity Laws, Joy's Law, polar field reversals, and the transport of higher latitude magnetic field elements toward the poles.'' (Hathaway, 2010). 
Black lines connect the times of a-f onto g.
Small colored circle show the direction of field line of (a-f) on g.
White lines show suggested paths of the dynamo-flux under the solar surface.}
\label{fig:6}
\end{figure}

Figures~\ref{fig:6} a) through f) depict the process of the polar flux reversal.
This is for a single hemisphere only, displaying the formation of new lobes, the displacement of old lobes, and a process by which the structure reverses polarity. 
This half-cycle shows the unwrapping of flux from around what becomes the new upward-directed lobe. 
The flux becomes the new (flipped) polar flux. 
The steps of the process are the following:

Figure~\ref{fig:6} a) The MECH state near solar minimum. 
(The pattern of the radial field at the solar surface is seen at 1997 in the lower Fig.~\ref{fig:6} g.)

Figure~\ref{fig:6} b) Resistive diffusion moves the boundaries (1997.3).

Figure~\ref{fig:6} c) The upper boundary of the GMS becomes too close to the surface and the plasma pressure outside the GMS becomes too low to support the boundary pressure required for equilibrium. 
Part of the GMS then moves up and breaks through the surface, breaking the old lobes in two (1999).

Figure~\ref{fig:6} d) Meridional flow moves the polar side (of the break) to the polar region but in the equator side the flow is countered by tension in the polar flux resulting in a slow movement toward the equator. 
The higher latitudes of the new penetrations move to the poles where some (gray) are released from the Sun and some (red) becomes the new polar flux, anchored by wrapping the red lobe, which is deep enough to stay submerged. 
The lower latitudes of the new penetrations slowly move (allowing time for sunspots) to the equator releasing the new polar flux from the solar surface and releasing the old polar flux from the Sun.
The high latitude streaks to the poles and the coloration at low latitudes in Fig.~\ref{fig:6} g) are consistent with this description. 
The gray lobe breaking off and floating to the surface provides the thin sheets of magnetized plasma for prominences, which often appear to lift off the surface. 
The new polar flux is being dynamo driven and a new lobe starts growing from the helicity injection (2004).

Figure~\ref{fig:6} e) Old upper lobe has left allowing the new polar flux to connect with its counterpart in the southern hemisphere to form the new polar flux (2006).

Figure~\ref{fig:6} f) New MECH state near solar minimum (2009). 
The polar flux came from the red lobe.

The time of the 11-year cycle is the time it takes the upper edge of the red lobe in Fig.~\ref{fig:6} d) to reach the level at which the upper lobe is released for lack of equilibrium. 
The GMS remains stable, and the lobe leaves from insufficient mass above that supplies the pressure it to
keep it submerged. 
When the dynamo drive starts up on the new polar flux, which was in the self-organized part of the GMS, the pinch effect compresses this flux bringing the lambda up to the injector value and causes the bottom to move up by the amount it moves down in the rest of the cycle (see Fig.~\ref{fig:6} d). 
Similarly, the loss of the upper lobe causes the top to move down by the amount it moves up in the rest of the cycle.
The size of the red lobe may become the width to make these motions have equal time.

The plasma resistivity is estimated using Spitzer resistivity (Spitzer, 1962) for the plasma plus a constant-cross-section ($5.7 \times 10^{- 19}$~m$^{2}$) correction for electron collisions with neutrals (Kolesnikov and Obukhov-Denisov, 1962). 
The Saha equation is used to find the percentage ionized. 
Estimates of solar properties are from Reference (Christensen-Dalsgaard, 1996). 
The resistivity in $\Omega$m as a function of depth below the photosphere $x$ in meters ($x$=0 at the surface) is approximately:

\begin{equation}
\eta = \frac{60}{x + 8000}.
\label{eq:7}
\end{equation}

The time for the lobe to diffuse to the release height is of order 11 years fairly independently of the value of $B_{o}$. 
The conditions that the three interfaces must have the same period determines the size, the depth and the cycle period of the GMS. 
More data from the dynamo dynamics in the Sun are needed to validate this result. 
A contribution of this paper is to show what to look for in the dynamo and where to look for the dynamo.

\subsection{Results: flares, supergranules, sunspots, and CMEs}

Flares are likely generated by similar physics as are on display at the poles during minima. 
The discussion of current drive so far has been for the solar minimum time, as in Fig.~\ref{fig:6} a). 
When the Sun is active, as in Fig.~\ref{fig:6} 6 c), every field line that enters the Sun at a different latitude than it exits the Sun, will have current drive and power by the same mechanism. 
Fig.~\ref{fig:6} c) shows the active Sun with current penetration over the whole surface, while the quiet Sun has only penetrations more-or-less confined to the equatorial regions and the polar regions.
Flares are seen in the corona and the observed (Platten \emph{et al.,} 2014) distribution of flares over the solar cycle agrees with the flux penetration distribution of the model. 
Thus, where flux is penetrating the solar surface, the dynamo power is released to the solar atmosphere as a flare. 
When a dynamo driven flux tube first breaks through the solar surface a large amount of energy is released in the process of reaching a steady state condition.

The supergranules are probably spheromaks that are the orthogonal perturbations needed for the grossly stable sustainment of the GMS by the dynamo-driven current parallel to the polar flux. 
The growth and decay times ($\approx$1 day) of the supergranules are about the meridian Alfv\'{e}n time (assuming $p \approx B_{o}^{2} / 2 \mu_0$ and $T \approx 1$~eV) and they appear to be the signature of relaxation-instability. 
One argument in favor of the GMS producing the supergranules is their pair-correlated movement up to 600~Mm (Hirzberger, 2008). 
This correlation on the solar radius scale is almost certainly due to magnetic effects. 
During solar minimum the GMS is sustained by the dynamo-driven polar flux requiring an orthogonal perturbation. 
A spheromak (Jarboe,1994) in a tuna-can shaped boundary of equal radius and height has a $\lambda_{eq} = 5/h$, where $h$ is the height. 
This is smaller than that of the self-organized part of the GMS, which is $\lambda_\text{self} \approx 2 \pi / h$, where $h$ is the height of the magnetized plasma volume. 
As the spheromak increases in radius, $\lambda_{eq}$ approaches $\pi / h$. 
(It is within 3\% at a radius of $5 h$.) 
Similarly, a hole through the center does not change $\lambda_{eq}$ significantly and thermal convective flow can freely pass through these regions. 
This would explain the circular shape (Hirzberger, 2008) and flow through the supergranules.
The flow pattern in supergranules region (Duvall, 2014) not only allows the solar power to pass through the GMS but also couples the dynamo torque to the convective plasma, which distributes it to the large volume of the torsional oscillations.

The spheromak is a lowest eigenstate for the plasma volume and, therefore, the unsustained plasma relaxes to this state. 
However, it is incompatible with the injector geometry for sustainment and equilibrium and decays as fast as it is formed. 
Evidently, the energy that would be required to bend the polar flux around the spheromak so that it could
sustain the spheromak, prevents this from being the sustained MECH state. 
Therefore, the MECH state sustained by the dynamo-driven polar flux is the GMS described. 
However, the supergranulation has the ideal geometry for the required orthogonal perturbation. 
(For resistive MHD the perturbations must have a component of $\delta \vec{v \times \delta B}$ parallel to $\vec{B}$ to sustain a MECH state.)
Alternatively, the spheromak has some magnetic fields that are directed oppositely to the polar flux and some that are directed oppositely to the self-organized part of the GMS fields. 
When the oppositely-directed fields reconnect, the spheromak connects the polar flux to the flux of the self-organized part of the GMS and helicity will flow from the polar flux to the self-organized part of the GMS. 
This anti-current drive in the polar flux and current drive in the self-organized part of the GMS sustains the GMS.

\begin{figure}
\begin{center}
\includegraphics[width=\linewidth]{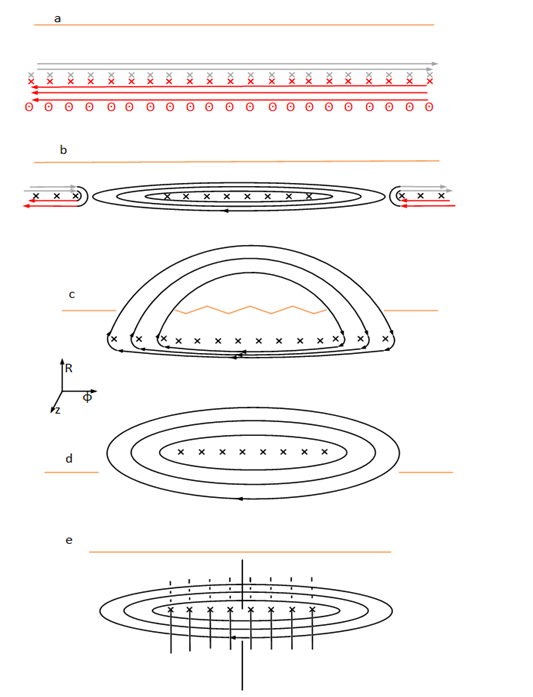}
\end{center}
\caption{Sketches of the possible formations and saturation of a sausage-like instability. 
The gold line below the frame label is the solar surface for that frame.
a) Stable GMS colors are as in Fig.~\ref{fig:6}. 
However, here the polar flux is into the page. 
b) Growth at the sun spot site due to helicity flowing to the site. 
Black is used to represent a mixture of gray and red. 
c) Sun spot formed. 
d) CME breaking through the solar surface. 
e) Tilting to form a supergrandule.}
\label{fig:7}
\end{figure}

Sunspots have too much flux to be formed from the local magnetic fields of the thin GMS. 
They are more likely simply formed in place by helicity injection from the solar dynamo and the rest of the GMS. 
Figure~\ref{fig:7} a) shows a cross section of a ribbon of the GMS. 
The ribbon is up to 50~Mm wide in the longitudinal direction and the same length as the dynamo in the latitudinal direction and a longitudinal cross section is shown. 
At the beginning of the cycle the polar flux can be thought of as the remaining part of the upper (gray) lobe that has partially escaped through the solar surface. 
The gray Xs represent this polar flux below the solar surface that will escape the Sun during this cycle. The red Xs represent the top of the lower lobe, which will become the new polar flux in the heliosphere. 
The ``out of the page'' symbols represent the lower part of the lower lobe, which will become the polar flux below the solar surface.

Sun spot formation can occur as follows: 
As the GMS moves to the surface, mass above the GMS becomes too small and is pushed away, and the GMS becomes locally thicker, causing a local lower of $\lambda_\text{self}$ which causes helicity to flow into the region. 
This region of the GMS expands by the inflow of flux and current that link the polar flux channel, producing a ``flat torus'' that is threaded by the polar flux (see Fig.~\ref{fig:7} b).
This is very similar to the well-known sausage instabilities.

The gold lines in Fig.~\ref{fig:7} represent the very resistive photosphere.
When this region bounds the flat torus, it no longer carries the current needed to confines the magnetic field of the upper thick area of the flat torus. 
(Faculae could be arcing where the flux conserver is failing, which is represented by a jagged solar surface.) 
The poloidal fields completely dissipate, and the toroidal field partially escapes but cannot leave because it links the polar flux. 
The trapped toroidal field enters and leaves the solar surface at the location of the
resistive gap as in Fig.~\ref{fig:7} c). 
If the dynamo current through this instability is disrupted, all of the flux linking this current becomes available to the sunspot. 
The linking flux is of order the polar flux, $\approx$$3 \times 10^{14}$~Wb, in agreement with the maximum of sunspot groups, $\approx$$2 \times 10^{14}$~Wb, (Priest, 2014). 
Thus, in this interpretation, the sunspots are formed at the footprints of the escaped toroidal field of the flat torus. 
The helicity and plasma that flow to the low-$\lambda$ region of the sunspot will be pushed downward by gravity giving the powerful converging and downward directed flows observed. (Zhao, 2001)
The angle of sunspots can vary because the angle of the resistive gap can vary. 
The toroidal field is not visible because it was stripped of its plasma as it passed through the insulating photosphere. 
The plasma escaping from the failing flux conserver may be observed as the Evershed clouds (Solana \emph{et al}., 2006). 
They have a speed like the sound speed of the surface plasma; they vent at the edge of the flux and in the direction of the flux there (horizontal) as might be expected. 
This venting would cause expansion cooling of the region viewed through the sunspot. 
The sunspot lasts until the old polar flux that is trapping it leaves the Sun, giving a large variation in the sunspot life time.

(Thus, the sunspot magnetic field tends to be perpendicular to the internal dynamo current. 
Joy's law (Hale \emph{et al}., 1919) for sunspots may occur because at first, near 30$^{\circ}$ latitude, the actual internal polar flux is at an angle to the meridian direction.)

If the flat torus also links some of the lower lobe's flux and is strong enough to pull this lobe flux and the polar flux out of the surface as perhaps in a CME (Fig.~\ref{fig:7} d)) then something like the transition from Fig~\ref{fig:6} b) to Fig.~\ref{fig:6} c) would occur, producing a flare. 
The magnetic fields in CMEs have this structure (Rakowski, 2011) and CME and flares tend to appear from the same event. 
CMEs have produced up to 15~nT of magnetic field at the earth (Owens, 2017) and multiplying by 1~au squared gives $3 \times 10^{14}$~Wb, indicating CMEs might be related to sunspots since their maximum flux are of the same order. 
If the flat torus is formed well below the surface it should tilt 90$^{\circ}$ and form a short lived spheromak of the supergrandules (Fig.~\ref{fig:7} e)). 
This activity may produce measurable radial fringe magnetic field at the surface perhaps related to the magnetic carpet (Priest, 2008). 
The magnetic field of the magnetic carpet appear to emerge from supergrandules.

\section{Discussion}

Self-organization properties of sustained magnetized plasma are applied to selected solar data to understand solar magnetic dynamics.
The solar magnetic activity and the torsional oscillations are well correlated. (Howe, 2009) 
The thickness of the solar dynamo can be estimated from the amplitude of the torsional oscillations it drives.
The result is a gradient scale length of about 100~km similar to that of the photosphere. 
Such a thin dynamo is powerful enough to drive all of the solar electro-magnetic activity. 
Anytime flux enters and leaves the sun at locations of different rotational rates, dynamo drive is produced. 
The acceleration of torsional oscillations agrees in detail with the radial magnetic field of magnetograms, which allow the identification of the location of dynamo activity throughout the solar cycle.

The dynamo accelerates the heliosphere until the back EMF equals the highest dynamo driven EMF, resulting in the rigid rotation of the heliosphere. 
This result is supported by the comparison of the rotation of the corona to magnetograms.

A thin stable magnetic equilibrium seems to be covering the solar surface just below the photosphere from latitudes of 15 degrees, or less, to at least 60 degrees. 
The magnetic field lines are expected to be parallel to the solar surface and rotate with distance from the
surface and to rotate $2\pi$ radians in $\approx$1~Mm. 
This is expected for the sustained configuration that has the minimum magnetic energy for the helicity content, in accordance to the well-established minimum energy principle of plasma self-organization. 
Resistive diffusion may help to push the magnetic fields to the surface and the Global Magnetic Structure (GMS) seems to lose $\pi$ radians every 11 years, causing the observed 180$^{\circ}$ flipping of the solar magnetic fields including the polar flux. 
Further evidence for this GMS and its loss is that solar prominences are made of thin sheets of magnetized plasma like the thin sheets lost by the GMS. 
The differences observed between solar maximum and solar minimum in the corona also agrees with the process presented. 
This evidence leaves little doubt the GMS is a sustained minimum energy state.

The roughly $\pi$ radians halves of the GMS are separated by the polar flux, which has a parallel current driven by the solar dynamo. 
Perturbations cause the driven current to drive the adjacent current of the GMS, sustaining the GMS. 
The magnetic perturbations may come from a transient state of lower energy than the GMS when it is not sustained but a higher energy if it were sustained, so it appears and decays, giving the necessary magnetic perturbations. 
The transient state would be the spheromak. 
The spheromak has the topology of, and is compatible with, the plasma flow pattern measured in the supergranules, which exist in the GMS region. 
The lifetimes of the supergranules agree with that expected for these transient states. 
Thus, it is quite plausible that transient spheromaks supply the perturbation for cross-field current drive. 
Alternately, the spheromaks may connect polar flux with the magnetic field of the rest of the GMS for the cross-field drive that would also sustains the GMS. 
The supergranules are the signature of the spheromaks.

The flow pattern in supergranules region allow the solar power to pass through the GMS and couples the dynamo torque to the convecting plasma, which distributes it to the large volume of the torsional oscillations.

For completeness, it is likely that when the mass above the GMS becomes too small to keep the GMS submerged that a sausage like bubble grows a flat torus with a toroidal flux that links some polar flux. 
If the polar flux is strong enough, it traps the toroidal flux as the flat torus breaks the surface and prevents the toroidal flux from leaving with the rest of the torus. 
Sunspots then appear where the toroidal flux leaves and enters the solar surface. 
If the toroidal flux of the flat torus is strong enough, it can pull the polar flux out of the surface as it leaves creating a CME and a flare. 
Deeper in the plasma, the flat torus may tilt to form a supergranule. 
All of these objects can dissipate quickly by leaving the solar surface.

\begin{acknowledgements}

The authors wish to thank many solar scientists for the highest quality data and Dr. Greg Kopp for help in estimating the dynamo power from the total solar irradiance data. 
This work supported by the U.S. Department of Energy Office of Science, Office of Fusion Energy Sciences under Awards No.\ DE-FG02-96ER54361 and No.\ DE-SC0016256.

\end{acknowledgements}

Amenomori, M., Bi, X.J., Chen, D. (2013). 
Probe of the solar magnetic field using the ``cosmic-ray shadow'' of the Sun. 
\emph{Astrophys. J.}, \textbf{762}, 131. doi:10.1088/0004-637X/762/2/131.

Alfvén, H. (1942). Existence of electromagnetic-hydrodynamic waves.
\emph{Nature} \textbf{150}, 405.

Babcock, H.W. (1953). The solar magnetograph. \emph{Astrophys. J.}
\textbf{118}, 387.

Babcock, H. W. (1961). The topology of the Sun's magnetic field and the
22-year cycle. \emph{Astrophys. J.} \textbf{133}, 572.

Birch, A.C., Braun, D.C., Leka, K.D., Barnes, G., Javornik, B. (2013).
Helioseismology of pre-emerging active regions. II. Average emergence
properties. \emph{Astrophys. J.}, \textbf{762}, 131.
doi:10.1088/0004-637X/762/2/13

Brandenburg, A. (2005). The case for a distributed solar dynamo shaped
by near-surface shear, \emph{Astrophys. J.}, \textbf{625}, 539.

Chandra, S., Om Vats, H., and Iyer, K. N. (2010). Differential rotation
measurement of soft X-ray corona, \emph{Mon. Not. R. Astron. Soc}.
\textbf{407,} 1108--1115 doi:10.1111/j.1365-2966.2010.16947.x

Charbonneau, P. (2010). Dynamo models of the solar cycle. \emph{Liv.
Rev. Solar Phys.} \textbf{7.1}, 1.

Charbonneau, P. (2014). Solar Dynamo Theory, Annu. Rev. Astron.
Astrophys. 52:251-90

Christensen-Dalsgaard, J., et al. (1996). Sound speed, etc., for Model
S. \emph{Science} \textbf{272}, 1286.

Christensen-Dalsgaard, J. (2002). Helioseismology. \emph{Rev. Mod.
Phys.} \textbf{74.4}, 1073.

Christensen-Dalsgaard, J., Thompson, M.J. (2007) Observational Results
and Issues Concerning the Tachocline. \emph{The Solar Tachocline:}
Cambridge University Press, 2007, 53.

Choudhuri, A. R., Schussler, M., Dikpati, M. (1995). The solar dynamo
with meridional circulation. \emph{Astron. Astrophys.} \textbf{303},
L29.

Choudhuri, A. R., Chatterjee, P., Jiang J. (2007). Predicting solar
cycle 24 with a solar dynamo model. \emph{Phys. Rev. Lett.}
\textbf{98.13}, 131103.

Cowling, T. G. (1934). The Magnetic Field of Sunspots. \emph{Month. Not.
Roy. Astron. Soc.} \textbf{94}, 39.

Dikpati, M., Charbonneau, P.: (1999). A Babcock-Leighton flux transport
dynamo with solar-like differential rotation, \emph{Astrophys. J.},
\textbf{518}, 508.

Duvall, T.L.Jr., Hanasoge, S.M., Chakraborty, S. (2014). Additional
evidence supporting a model of shallow, high-speed supergranulation.
\emph{Solar Phys.} \textbf{289}, 3421. doi: 10.1007/s11207-014-0537-3

Edenstrasser, J.W., Kassab, M.M.M. (1995). The Plasma Transport
Equations Derived by Multiple Timescale Expansion: An Application,
\emph{Phys. Plasmas} \textbf{2}, 1206

Finn, J.M., Antonsen T.M.Jr. (1985). Magnetic helicity: What is it and
what is it good for. \emph{Comm. Plas. Phys. Cont. Fus.} \textbf{9.3},
111.

Haber, D.A., \emph{et al.} (2002). Evolving submerged meridional
circulation cells within the upper convection zone revealed by
ring-diagram analysis. \emph{Astrophys. J.} \textbf{570.2}, 855.

Hale, G.E. (1908). On the probable existence of a magnetic field in
sun-spots. \emph{Astrophys. J.} \textbf{28}, 315.

Hale, G. E., Ellerman, F., Nicholson, S.B., Joy, A.H. (1919). The
magnetic polarity of sun-spots. \emph{Astrophys. J.} \textbf{49}, 153.

Hathaway, D.H., et al. (1996). GONG observations of solar surface flows.
\emph{Science, New Series}, \textbf{272.5266}, 1306.

Hathaway, D.H. (2010). The solar cycle, \emph{Liv. Rev. Solar Phys}.,
\textbf{7}, 1.

Hindman, B.W., et al. (2004). Comparison of solar subsurface flows
assessed by ring and time-distance analyses. \emph{Astrophys. J.}
\textbf{613}, 1253.

Hirzberger, J., Gizon, L., Solanki, S.K., Duvall, T.L. Jr. (2008)
Structure and evolution of supergranulation from local helioseismology,
\emph{Solar Phys}. \textbf{251,} 417.

A. C. Hossack et al. (2017). Derivation of dynamo current drive in a
closed-current volume and stable current sustainment in the HIT-SI
experiment. \emph{Phys. Plasmas} \textbf{24}(3), 020702.

Howe, R. (2009). Solar Interior Rotation and its Variation \emph{Living
Rev. Solar Phys}., \textbf{6}, 1.

B. Hudson, R. D. Wood, H. S. McLean, E. B. Hooper, D. N. Hill, J.
Jayakumar, J. Moller, D. Montez, C. et al. (2008). Energy confinement
and magnetic field generation in the SSPX spheromak. \emph{Phys Plasmas}
\textbf{15}, 056112

Jarboe, T.R. (1994). Review of Spheromak Research. \emph{Plasma Phys.
Control Fusion} \textbf{36}, 945.

Jarboe, T.R., Victor, B.S., Nelson, B.A., Hansen, C.J., Akcay, C.,
Ennis, D.A., Hicks, N.K., Hossack, A.C., Marklin, G.J., Smith, R.J.
(2012). Imposed-dynamo current drive. \emph{Nucl. Fusion}~\textbf{52},
083017.

Jarboe, T. R., Sutherland, D. A and Nelson, B.A. (2015). A mechanism for
the dynamo terms to sustain closed-flux current, including helicity
balance, by driving current which crosses the magnetic field.
\emph{Phys. Plasmas,} \textbf{22}.

Jiang, J., Cameron, R.H., Schmitt, D., Schüssler, M. (2011).The solar
magnetic field since 1700 II. Physical reconstruction of total, polar,
and open flux. \emph{Astron. Astrophys.} \textbf{528}, A83. doi:
10.1051/0004-6361/201016168

Klimchuk, J.A. (2006). On solving the coronal heating problem.
\emph{Solar Phys.} \textbf{234\emph{, }}41.

Kolesnikov, V.N., Obukhov-Denisov, V.V. (1962). The effective cross
section for elastic scattering of slow electrons by hydrogen atoms.
\emph{Soviet Physics Jetp} \textbf{15.}

Kopp, G. (2016). Magnitudes and timescales of TSI variability, \emph{J.
Space Weath. Space Clim.}, A30.

Landin N. R., Mendes, L. T. S. and Vaz L. P. R. (2010). Theoretical
values of convective turnover times and Rossby numbers for solar-like,
pre-main sequence stars. Astronomy \& Astrophysics 510, A46 DOI:
10.1051/0004-6361/200913015

Larmor, J. (1919). How could a rotating body such as the Sun become a
magnet? \emph{Rep. Br. Assoc. Adv. Sci.} \textbf{A}, 159.

Leighton, R.B. (1969). A magneto-kinematic model of the solar cycle.
\emph{Astrophys. J.} \textbf{156}, 1.

Maunder, E.W., Maunder, A.S.D. (1904). Note on the distribution of
sun-spots in heliographic latitude, 1986-1902. \emph{Month. Not. Roy.
Astron. Soc.} \textbf{64}, 747.

Moffat, H.K. (1978). Magnetic Field Generation in Electrically
Conducting Fluids. Cambridge University Press, New York, 1978.

Owens, M. J., Lockwood, M \& Barnard, L. A. (2017). Coronal mass
ejections are not coherent magnetohydrodynamic structures. Scientific
Reports \textbf{7}: 4152\emph{.} DOI:10.1038/s41598-017-04546-3

Parker, E.N. (1955). Hydromagnetic dynamo models. \emph{Astrophys. J.}
\textbf{122}, 293.

Parker, E.N. (1963). The Solar-Flare Phenomenon and the Theory of
Reconnection and Annihilation of Magnetic Fields. \emph{Astrophy. J.
Supp. Ser.} \textbf{8}, 177.

Parker, E.N.
(1993).~\href{http://adsabs.harvard.edu/full/1993ApJ...408..707P}{A
solar dynamo surface wave at the interface between convection and
nonuniform rotation}''. \emph{The Astrophysical Journal}, \textbf{408},
707.

Parker, E.N. (2008). Solar magnetism: the state of our knowledge and
ignorance. In: Thompson, M.J, Balogh, A. et al. (eds.) The Origin and
Dynamics of Solar Magnetism, Springer, New York, 2008. 15.

Platten, S.J. Parnell, C.E., Haynes, A.L., Priest, E.R., Mackay, D.H.
(2014).The Solar Cycle Variation of Topological Structures in the Global
Solar Corona. \emph{Astron. Astrophys.} \textbf{565}, A44.

Priest, E.R. Heyvaerts, J.F., Title, A.M. (2002). A Flux Tube Tectonics
Model for solar coronal heating driven by the magnetic carpet.
\emph{Astrophys. J}., \textbf{576}, 533.

Priest, E. (2014). Magnetohydrodynamics of the Sun, Cambridge University
Press.

R\"{a}dler, K-H. (2014). Mean-field dynamos: The old concept and some recent
developments. \emph{Astronomische Nachrichten} \textbf{335}, 459.

Rakowski, C.E., Laming, J.M.,and Lyutikov, M. (2011) In situ heating of
the 2007 may 19 cme ejecta detected by stereo/plastic and ace. \emph{The
Astrophys J.} \textbf{730}(30). 8.

Savage, D., Steigerwald, B., Title, A.M. (1997). \emph{Nasa Press
Release No: N97-147 (HQ 97-256).}

Schnack, D.D. (2009). \emph{Lectures in Magnetohydrodynamics: With an
appendix on Extended MHD}, Springer, Berlin Heidelberg. \textbf{262}.
doi:10.1007/ 978-3-642-00688-3.

Smith, E.J. (2001). The heliospheric current sheet. \emph{J. Geophys.
Research} \textbf{106}, 15819.

Solana, D. C., \emph{et al.} (2006). Evershed clouds as precursors of
moving magnetic features around sunspots. \emph{Astrophys. J. Lett}.
\textbf{649}, L41.

Spitzer, L.Jr. (1962). Physics of fully ionized gas. Interscience
(Wiley), New York, 1962.

Spruit, H.C., 2003, ``Origin of the torsional oscillation pattern of
solar rotation'', Solar Phys., 213, 1--21.

Steenbeck, M., Krause, F., Rädler, K.-H. (1966). Berechnung der
mittleren Lorentz-Feldstärke $ \overrightarrow{v \times b}$  für ein
elektrisch leitendes Medium in turbulenter, durch Coriolis-Kräfte
beeinflußter Bewegung. \emph{Z. Naturforsch.} \textbf{21a}, 369.

Stenflo, J.O. (2015). History of solar magnetic fields since George
Ellery Hale. \emph{Space Sci. Rev.}, \textbf{1}.

Taylor, J.B.Jr. (1986). Relaxation and magnetic reconnection in plasmas.
\emph{Rev. Mod. Phys}. \textbf{58}, 741.

Tobias, S.M. (2002). The solar dynamo. \emph{Phil. Trans. Roy. Soc.
London A: Math. Phys. Eng. Sci.} \textbf{360,} 2741.

Vernazza, J.E., Avrett, E.H., Loeser, R. (1981). Structure of the solar
chromosphere. III-Models of the EUV brightness components of the
quiet-sun. \emph{Astrophys. J. Supp. Ser.}, \textbf{45}, 635.

Victor, B.S., Jarboe, T.R., Hansen, C.J., Akcay, C., Morgan, K.D.,
Hossack, A.C., Nelson, B.A. (2014). Sustained spheromaks with ideal n =
1 kink stability and pressure confinement. \emph{Phys. Plasmas},
\textbf{21}, 082504.

Woltjer, L. (1958). A theorem on force-free magnetic fields. \emph{Proc.
Natl. Acad. Sci.} \textbf{44}, 489.

Yamada, M., Kulsrud, R., Hantao, J. (2010). Magnetic Reconnection,
\emph{Rev. of Mod, Phys.}, \textbf{82}, 603.

Zhao, J, Alexander G., Kosovichev, A. G., and Duvall, T. L. Jr. (2001). Investigation of mass flows beneath a sunspot by time-distance helioseismology, \emph{The Astrophysical Journal}, \textbf{557}, 384.

\end{document}